Nonlinear damping in mechanical resonators based on graphene and carbon nanotubes

A. Eichler[1,*], J. Moser[1,*], J. Chaste[1], M. Zdrojek[1], I. Wilson-Rae[2], A. Bachtold[1]

[1] CIN2(ICN-CSIC), Catalan Institute of Nanotechnology, Campus de la UAB 08193 Bellaterra (Barcelona) Spain

[2] Technische Universität München, 85748 Garching, Germany

[*] These authors contributed equally to this work.

**The theory of damping finds its roots in Newton's Principia [1] and has been exhaustively tested in objects as disparate as the Foucault pendulum, mirrors used in gravitational-wave detectors, and submicron mechanical resonators. Owing to recent advances in nanotechnology it is now possible to explore damping in systems with transverse dimensions on the atomic scale. Here, we study the damping of mechanical resonators based on a carbon nanotube [2-11] as well as on a graphene sheet [12-15], the ultimate one and two-dimensional nanoelectromechanical systems (NEMS). The damping is found to strongly depend on the amplitude of the motion; it is well described by a nonlinear force $\eta x^2 \dot{x}$ (with $x$ the deflection and $\dot{x}$ its time derivative). This is in stark contrast to the linear damping paradigm valid for larger mechanical resonators. Besides, we exploit the nonlinear nature of the damping to improve the figure of merit of nanotube/graphene resonators.**

Damping is a key phenomenon in NEMS resonators. Not only does it impact the resonator dynamics (namely its motional amplitude and velocity), it also governs the performance of the resonator in various scientific and technological applications. These include studies of the quantum-to-classical transition [16], the cooling efficiency [17], the mass resolution [18], and the force sensitivity [19]. Damping has been successfully described by the linear damping force $\gamma \dot{x}$ for all the mechanical resonators studied so far. Remarkably, this picture holds for resonators whose dimensions span many orders of magnitude down to a few tens of nanometers. Reducing dimensions to the atomic scale using graphene and nanotube resonators, we show that a simple linear damping scenario ceases to be valid. To demonstrate this, we provide a detailed experimental study showing that the quality factor strongly varies with the driving force [12] and we analyze this behaviour in light of the nonlinear damping theory [20],

We perform measurements on graphene/nanotube resonators (Fig. 1a,b) at low temperature and in high vacuum, using a dilution refrigerator with a base temperature of 90 mK. The resonator is actuated electrostatically by applying an oscillating voltage $V^{AC}$ at frequency $f$ between the resonator and a gate electrode (Fig. 1c). The motion is detected using the frequency-modulation (FM) mixing technique where the resonator acts as a frequency mixer to deliver a mixing current $I_{mix} \propto \left| \partial/\partial f \, \text{Re}[x_0] \right|$ (with Re[$x_0$] the real part of the motional amplitude $x_0$) [10]. The mixing current as a function of driving frequency $f$ has a characteristic resonance line-shape that allows us to extract the mechanical quality factor $Q$ in a simple manner (Fig. 1d): the resonance peak at frequency $f_0$ is flanked by two minima whose separation is the resonance width $\Delta f = f_0 / Q$ for a linear harmonic oscillator. This simple relation is expected to break down in the presence of nonlinearities (see Fig. 1e) but an analogous expression is recovered in the limit of strong nonlinear damping, as discussed below (and in the Supplementary Information, section D).

To show that nonlinear damping in graphene and nanotube NEMSs is a robust phenomenon, we study three types of mechanical resonators: (i) nanotube under tensile stress, (ii) nanotube with slack, and (iii) graphene sheet under tensile stress. We estimate the built-in stress in each of these devices by measuring their basic mechanical properties. As an example, Fig. 2a displays the dependence of the resonance frequency on gate voltage $V_g^{DC}$ for a nanotube resonator. The convex parabola has an electrostatic origin [21, 22] and indicates that the nanotube is under tensile stress (schematic diagram of Fig. 2a) [14,15].

We arrive at the central result of the paper. Fig. 2b shows the resonant response of the stressed nanotube resonator for three different driving forces (these scale linearly with $V^{AC}$). As we increase the driving force, the resonance frequency shifts towards higher values and, simultaneously, the resonance peak broadens (see bars below the resonances). Both these effects are also displayed in Fig. 2c,d. In these measurements, care is taken to avoid driving $V^{AC}$ above $k_B T/e$ in order to prevent electronic nonlinear effects or local heating (here $k_B$ is the Boltzmann constant, $T$ the temperature, and $e$ the electron charge). While the resonance shift is a known behavior (see below), the resonance broadening is a novel phenomenon. In larger resonators, the resonance width is indeed independent of the driving force ($\Delta f = f_0/Q = \gamma/2\pi m$ where $m$ is the mass of the resonator).

The same measurement is performed on the nanotube with slack (schematic of Fig. 2e) and on the graphene sheet under tensile stress (schematic of Fig. 3a). The resonance broadening is observed in all three types of resonators (Fig. 2c, Fig. 2e, Fig. 3a) and even at room temperature (Supplementary Information, Fig. S10). This validates the robustness of the effect and confirms early optical measurements on graphene [12] showing similar behaviour. The resonance broadening does not stem from the coupling between electrons and mechanical vibrations [8,9] because the effect is not associated to Coulomb blockade and $V^{AC}$ is kept below $k_BT/e$ (more discussions in Supplementary Information, section J). The resonance shift shows different behaviors: It is significant for the resonators under tensile stress (Fig. 2d, Fig. 3b), yet it is negligible (Fig. 2f) and sometimes even negative (Supplementary Information, Fig. S10) for nanotube resonators with slack. Further discussion, as well as additional electrical and mechanical characterizations, can be found in the Supplementary Information (sections E-G).

Upon further increasing $V^{AC}$, we observe a hysteretic response for the graphene resonator but not for any of the two nanotube resonators. In the case of the graphene resonator, the resonance lineshape differs depending on whether the driving frequency is swept upwards or downwards (Fig. 3c,d). The hysteresis is intimately related to the resonance shift [22-24]. They both originate from the so-called Duffing force $\alpha x^3$ [20]. The latter contributes to the restoring force, which makes the resonator stiffer (for $\alpha > 0$) and increases the resonance frequency. For sufficiently large driving forces, the motional amplitude as a function of the driving frequency *f* develops an asymmetry (black curve in the schematic of Fig. 3c). This results in bistability and hysteresis for certain intervals in *f* (red curves in the schematics of Fig. 3c,d). In this context, the absence of a hysteresis in some of our devices is intriguing.

We now show that both the broadening of the resonance and the occasional absence of the hysteresis can be understood within a single generalized nonlinear framework. In addition to the Duffing nonlinearity $\alpha x^3$, the other relevant higher-order term in the Newton equation for a harmonic oscillator is the nonlinear damping term $\eta x^2 \dot{x}$ [20,25],

$$m\ddot{x} = -kx - \gamma\dot{x} - \alpha x^3 - \eta x^2 \dot{x} + F_{drive}\cos(2\pi ft). \qquad (1)$$

Here $F_{drive}$ is the driving force amplitude and *k* the spring constant. Equation 1 provides a general treatment of nonlinear resonators, in the sense that, in the limit of weak damping and weak anharmonicity, additional terms of second and third order ($x^2$, $x\dot{x}$, $\dot{x}^2$, $x\dot{x}^2$, $\dot{x}^3$) merely lead to a renormalization of $\alpha$ and $\eta$ [20].

Dissipation is described by the terms $\gamma\dot{x}$ and $\eta x^2 \dot{x}$. The latter term is special as it accounts for a dissipation mechanism that becomes important at large motional amplitude. When $\gamma\dot{x}$ is dominant over $\eta x^2 \dot{x}$, which is the case for larger mechanical resonators, the resonance width is independent of the driving force and is given by $\Delta f = \gamma/2\pi m$. In the other limit, when the $\gamma\dot{x}$ term can be neglected, we obtain (see Supplementary Information, section D)

$$\Delta f = 0.032 m^{-1} \eta^{1/3} f_0^{-2/3} F_{drive}^{2/3} \qquad (2)$$

so that $\Delta f \propto (V^{AC})^{2/3}$. This dependence is in good agreement with the experimental data (red lines in Fig. 2c,e and 3a) and is used to extract $\eta$. ($\Delta f$ tends to saturate at low $V^{AC}$ for some devices, which may signal that linear damping begins to play a role; see Fig. 3a.) The shift of the resonance frequency as a function of $V^{AC}$ is determined from the maximum of $I_{mix}$ and is then compared to the steady-state solution of equation (1) using $\alpha$ as fit parameter (see Supplementary Information, section H). The agreement between theory and experiment is satisfactory (red line in Fig. 2d, f and 3b).

The occasional absence of the hysteresis is a direct consequence of nonlinear damping and can be predicted from the ratio between $\alpha$ and $\eta$. When $\eta/\alpha > \sqrt{3}/2\pi f_0$, the nonlinear damping is strong enough to keep the broadening of the resonance always comparable to or larger than its shift and precludes hysteresis for all driving forces [20]. Using the values of $\alpha$ and $\eta$ obtained from the aforementioned fitting, we find that $\eta/\alpha$ is larger than $\sqrt{3}/2\pi f_0$ for the two nanotube resonators in Fig. 2. Thus no hysteresis is expected, in agreement with the experiment. In contrast, we obtain $\eta/\alpha < \sqrt{3}/2\pi f_0$ for the graphene resonator in Fig. 3 and the predicted hysteresis is indeed observed.

The physical origin of nonlinear damping is a subtle problem that has been thus far underappreciated. A possible explanation is that it stems from the concerted effect of (1) a standard dissipation channel, which alone would lead to purely linear damping, and (2) geometrical nonlinearity, which can arise from the elongation of a doubly-clamped resonator upon deflection (see section E of Supplementary Information for experimental evidences of the geometrical nonlinearity effect). Such a scenario has been analyzed for a dissipation mechanism described by a phenomenological viscoelastic model [25] yielding the relation $\eta = 4\gamma/r^2$ for a rod under tensile stress. This leads to $\eta < 260$ kg·m$^{-2}$s$^{-1}$ from Fig. 2c (where we can estimate an upper bound for $\gamma$ using $2\pi m \cdot \min(\Delta f) = 2.5 \cdot 10^{-16}$ kg·s$^{-1}$). This does not compare well with the value obtained from the fit ($\eta = 10^4$ kg·m$^{-2}$s$^{-1}$), which suggests that the underlying physics is different. The viscoelastic model assumes that the dissipation is internal to the resonator, so the observed nonlinear damping could be associated to a dissipation channel exterior to the resonator, for example clamping losses (phonon tunneling [26]). Alternatively, geometric nonlinearity may not play any role and the nonlinearity of the damping may be germane to the dissipation mechanism itself, e.g. friction associated to the sliding between the nanotube/graphene and the metal electrode. Another possible contribution could stem from the nonlinearities in phonon-phonon interactions. However, theoretical analyses of nonlinear damping are scarce, possibly because it was so far deemed irrelevant, and certainly more work is required.

One may wonder whether the relationship between $\Delta f$ and $Q$ remains meaningful when the damping is strongly nonlinear. Provided that $\Delta f$ and the resonance shift are much smaller than $f_0$, the standard definition of $Q$ in terms of the free-ringdown is still warranted and reads $Q = 2\pi E / \Delta E$ where $\Delta E$ is the mechanical energy lost over one oscillation period and $E$ is the corresponding time-averaged stored energy. Interestingly, equation (1) yields a quality factor that depends on $|x_0|$, the modulus of the slowly decaying free oscillation amplitude, and is given by $Q = 8\pi f_0 m / \eta \cdot |x_0|^{-2}$ when $\gamma$ is neglected. To make the connection with a driven resonator, we take $|x_0|$ as the maximum amplitude of the response and find that Q satisfies

$$Q = 1.09 f_0 / \Delta f \tag{3}$$

(see Supplementary Information, section D). This simple relation is all the more surprising because it is very close to that of a simple harmonic oscillator.

It follows that our control over the resonance width allows us to improve the mechanical quality factor. In order to achieve larger *Q*-factors, we simply lower the driving force until the motion becomes barely detectable. For this, it is convenient to select the value of $V_g^{DC}$ for which the detection signal is largest. In so doing, we measure a quality factor of 100,000 for a graphene resonator at 90 mK (Fig. 4a). This is the largest *Q* ever reported in a graphene resonator.

Larger quality factors enable better force sensing. Fig. 4b shows the resonances of a nanotube at very low driving forces (with $V^{AC}$ as low as 200 nV). Using $C' = 5.2 \cdot 10^{-12}$ F/m (see Supplementary Information, section E), $V_g^{DC} = 2.49$ V, and the 1 Hz measurement bandwidth, we obtain a force sensitivity of 2.5 aN·Hz$^{-1/2}$ (here the force is $C'V_g^{DC}V^{AC}$, with *C'* the derivative of the gate-resonator capacitance with respect to *x*). This is within a factor of five of the best sensitivities reported using microfabricated resonators operating at their ultimate limit set by thermal vibrations [27, 28]. As there is room to optimize the detection scheme, the sensitivity of nanotube/graphene resonators can in principle be further enhanced.

In conclusion, the strong nonlinear damping constitutes a new regime for mechanical resonators. It is a robust phenomenon, as it is observed in two distinct systems (graphene and nanotube resonators) and is independent of the built-in tension (tensile stress or slack). Our finding entails that many predictions concerning quantum and sensing experiments ought to be revisited when applied to nanotube/graphene resonators, since they are based on the linear damping paradigm (e.g. [16-19]). The nonlinear damping and the associated ability to tune the quality factor hold promise for various scientific and technological applications.

**Methods**

We employ three different strategies to fabricate our resonators. In one approach, we grow nanotubes via chemical-vapor deposition on an oxidized silicon wafer. Nanotubes are contacted to metal electrodes by electron-beam lithography and are suspended in a wet etching step. Alternatively, we grow the nanotube in the last fabrication step over a predefined trench separating two electrodes [9]. Lastly, we fabricate graphene resonators by depositing a single graphene layer onto an oxidized silicon wafer using the adhesive tape technique [29]. We contact the graphene sheet to metal electrodes and suspend it by etching the silicon oxide [30]. See also Supplementary Information.

**References**

1. Newton, I., Principia, Book II (1687).

2. Sazonova, V., Yaish, Y., Üstünel, H., Roundy, D., Arias, T. A., & McEuen, P. L. A tunable carbon nanotube electromechnaical oscillator. *Nature* **431,** 284-287 (2004).

3. Garcia-Sanchez, D., San Paulo, A., Esplandiu, M. J., Perez-Murano, F., Forro, L., Aguasca, A., & Bachtold, A. Mechanical detection of carbon nanotube resonator vibrations. *Phys. Rev. Lett.* **99,** 085501 (2007).

4. Lassagne, B., Garcia-Sanchez, D., Aguasca, A., & Bachtold, A. Ultrasensitive Mass sensing with a nanotube electromechanical resonator. *Nano Lett.* **8,** 3735-3738 (2008).

5. Chiu, H.-Y., Hung, P., Postma, H. W. Ch., & Bockrath, M. Atomic-scale mass sensing using carbon nanotube resonators. *Nano Letters* **8,** 4342-4346 (2008).

6. Jensen, K., Kim, K., & Zettl, A. An atomic-resolution nanomechanical mass sensor. *Nature Nanotech.* **3,** 533-537 (2008).

7. Hüttel, A. K., Steele, G. A., Witkamp, B., Poot, M., Kouwenhoven, L. P., & van der Zant, H. S. J. Carbon nanotubes as ultrahigh quality factor mechanical resonators. *Nano Letters* **9,** 2547-2552 (2009).

8. Lassagne, B, Tarakanov, Y., Kinaret, J., Garcia-Sanchez, D., & Bachtold, A. Coupling mechanics to charge transport in carbon nanotube mechanical resonators. *Science* **325,** 1107-1110 (2009).

9. Steele, G. A., Hüttel, A. K., Witkamp, B., Poot, M., Meerwaldt, H. B., Kouwenhoven, L. P., & van der Zant, H. S. J. Strong coupling between single-electron tunneling and nanomechanical motion. *Science* **325,** 1103-1107 (2009).

10. Gouttenoire, V., Barois, T., Perisanu, S., Leclercq, J.-L., Purcell, S. T., Vincent, P., & Ayari, A. Digital and FM demodulation of a doubly clamped single-walled carbon-nanotube oscillator: towards a nanotube cell phone. *Small* **6,** 1060-1065 (2010).

11. Wang, Z., Wei, J., Morse, P., Dash, J. G., Vilches, O. E., & Cobden, D. H. Phase transitions of adsorbed atoms on the surface of a carbon nanotube. *Science* **327,** 552-555 (2010).

12. Bunch, J. S., van der Zande, A. M., Verbridge, S. S., Frank, I. W., Tanenbaum, D. M., Parpia, J. M., Craighead, H. G., & McEuen, P. L. Electromechanical resonators from graphene sheets. *Science* **315,** 490-493 (2007).

13. Garcia-Sanchez, D., van der Zande, A. M., San Paulo, A., Lassagne, B., McEuen, P. L., & Bachtold, A. Imaging mechanical vibrations in suspended graphene sheets. *Nano Lett.* **8,** 1399-1403 (2008).

14. Chen, C., Rosenblatt, S., Bolotin, K. I., Kalb, W., Kim, P., Kymissis, I., Stormer, H. L., Heinz, T. F., & Hone, J. Performance of monolayer graphene nanomechanical resonators with electrical readout. *Nature Nanotech.* **4,** 861-867 (2009).

15. Singh, V., Sengupta, S., Solanki, H. S., Dhall, R., Allain, A., Dhara, S., Pant, P., & Deshmukh, M. M. Probing thermal expansion of graphene and modal dispersion at low-temperature using graphene nanoelectromechanical systems resonators. *Nanotechnology* **21,** 165204 (2010).

16. Katz, I., Retzker, A., Straub, R., & Lifshitz, R. Signatures for a classical to quantum transition of a driven nonlinear nanomechanical resonator. *Phys. Rev. Lett.* **99**, 040404 (2007).

17. Kippenberg, T. J., & Vahala, K. J. Cavity optomechanics: Back-action at the mesoscale. *Science* **321,** 1172-1176 (2008).

18. Ekinci, K.L., Yang, Y.T., & Roukes, M. L. Ultimate limits of inertial mass sensing based upon nanoelectromechanical systems. *J. Appl. Phys.* **95**, 2682-2689 (2004).

19. Cleland, A. N., & Roukes, M. L. Noise processes in nanomechanical resonators. *J. Appl. Phys.* **92**, 2758-2769 (2002).

20. Lifshitz, R., & Cross, M. C. *Reviews of Nonlinear Dynamics and Complexity* (Wiley-VCH, New York, 2008, Vol. 1) [www.tau.ac.il/~ronlif/pubs/RNDC1-1-2008-preprint.pdf].

21. Howe, R. T. & Muller R. S. Resonant- microbridge vapor sensor. *IEEE Trans. on electron devices* **ED-33,** 499-506 (1986).

22. Kozinsky, I., Postma, H. W. C., Bargatin, I., & Roukes, M. L. Tuning nonlinearity, dynamic range, and frequency of nanomechanical resonators. *Appl. Phys. Lett.* **88,** 253101 (2006).

23. Aldridge, J. S. & Cleland, A. N. Noise-enabled precision measurements of a Duffing nanomechanical resonator. *Phys. Rev. Lett.* **94,** 156403 (2005).

24. Unterreithmeier, Q. P., Faust, T., & Kotthaus, J. P. Nonlinear switching dynamics in a nanomechanical resonator. *Phys. Rev. B* **81,** 241405R (2010)

25. Zaitsev, S., Shtempluck, O., Buks, E., & Gottlieb, O. *arXiv:*09110833v2.

26. Wilson-Rae, I. Intrinsic dissipation in nanomechanical resonators due to phonon tunnelling. *Phys. Rev. B* **77**, 245418 (2008).

27. Mamin, H. J. & Rugar, D. Sub-attonewton force detection at milikelvin temperatures. *Appl. Phys. Lett.* **79,** 3358 (2001).

28. Teufel, J. D., Donner, T., Castellanos-Beltran, M. A., Harlow, J. W., & Lehnert, K. W. Nanomechanical motion measured with an impresicion below that at the standard quantum limit. *Nature Nanotech.* **4,** 820-823 (2009)

29. Novoselov, K. S., Jiang, D., Schedin, F., Booth, T. J., Khotkevich, V. V., Morozov, S. V., & Geim, A. K. Two-dimensional atomic crystals. *Proc. Natl. Acad. Sci. U.S.A.* **102,** 10451-10453 (2005).

30. Moser, J. & Bachtold, A. Fabrication of large addition energy quantum dots in graphene. *Appl. Phys. Lett.* **95,** 173506 (2009).

**Acknowledgements**
We acknowledge support from the European Union (RODIN, FP7), the Spanish ministry (FIS2009-11284), the Catalan government (AGAUR, SGR), the Swiss National Science Foundation (PBBSP2-130945), and a Marie Curie grant (271938). IWR acknowledges financial support via the Nanosystems Initiative Munich. We thank Brian Thibeault (Santa Barbara) for help in fabrication and P. Gambardella, S. Roche, and S. Valenzuela for a critical reading of the manuscript.

**Author Contributions**
A.E., J.M., and M.Z. fabricated the devices. J.M. and A.E. developed the measurement setup and performed the measurements, J.C. and A.B. provided measurement support. A.E., J.M., A.B., and I.W.-R. analyzed the data. I.W.-R. established equations (2) and (3). A.B. conceived and designed the experiment. All authors contributed to writing the manuscript.

**Author information**
Correspondence and requests for materials should be addressed to A.B. (adrian.bachtold@cin2.es).

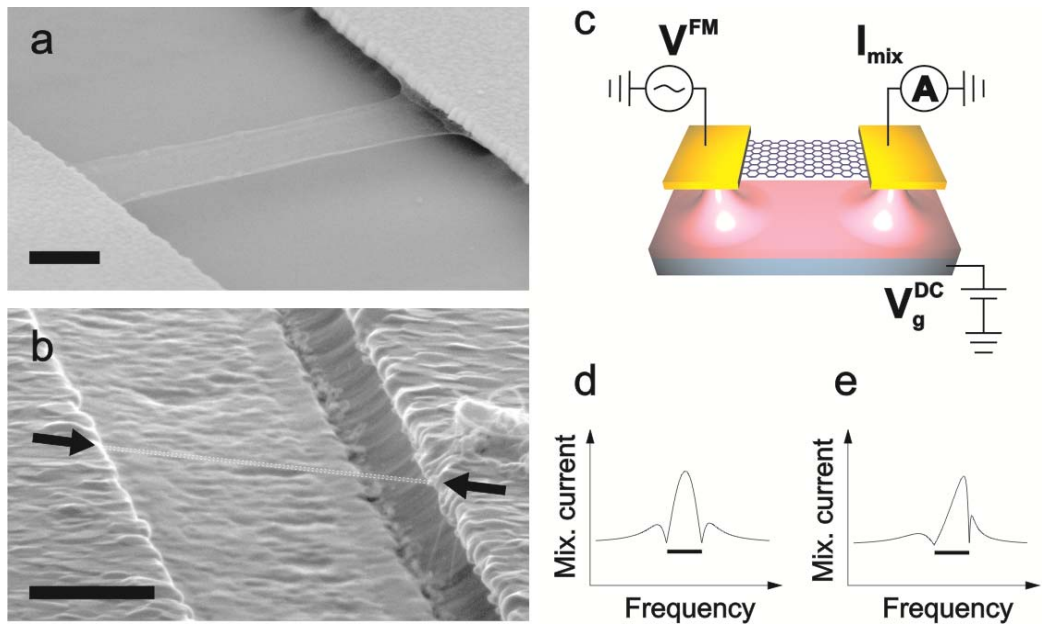

**Figure 1 | Devices and measurement setup. a,** Scanning electron microscopy image of a suspended single-layer graphene sheet with Au contacts. Scale bar, 500 nm. **b,** Scanning electron micrograph of a nanotube grown by chemical vapor deposition over a prefabricated trench between two W/Pt contacts. The nanotube is marked by black arrows and white dotted lines. Scale bar, 500 nm. **c,** Schematic of the actuation/detection setup. A frequency modulated voltage $V^{FM} = V^{AC} \cos[2\pi f t + f_\Delta / f_L \cdot \sin(2\pi f_L t)]$ is applied to the device. The motion is detected by measuring the mixing current at $f_L$. **d,** Schematic of the frequency response of the mixing current. The separation between the two minima (black bar) corresponds to the resonance width $\Delta f$, equal to $f_0 / Q$. **e,** The resonance peak becomes asymmetric in the presence of the Duffing nonlinearity.

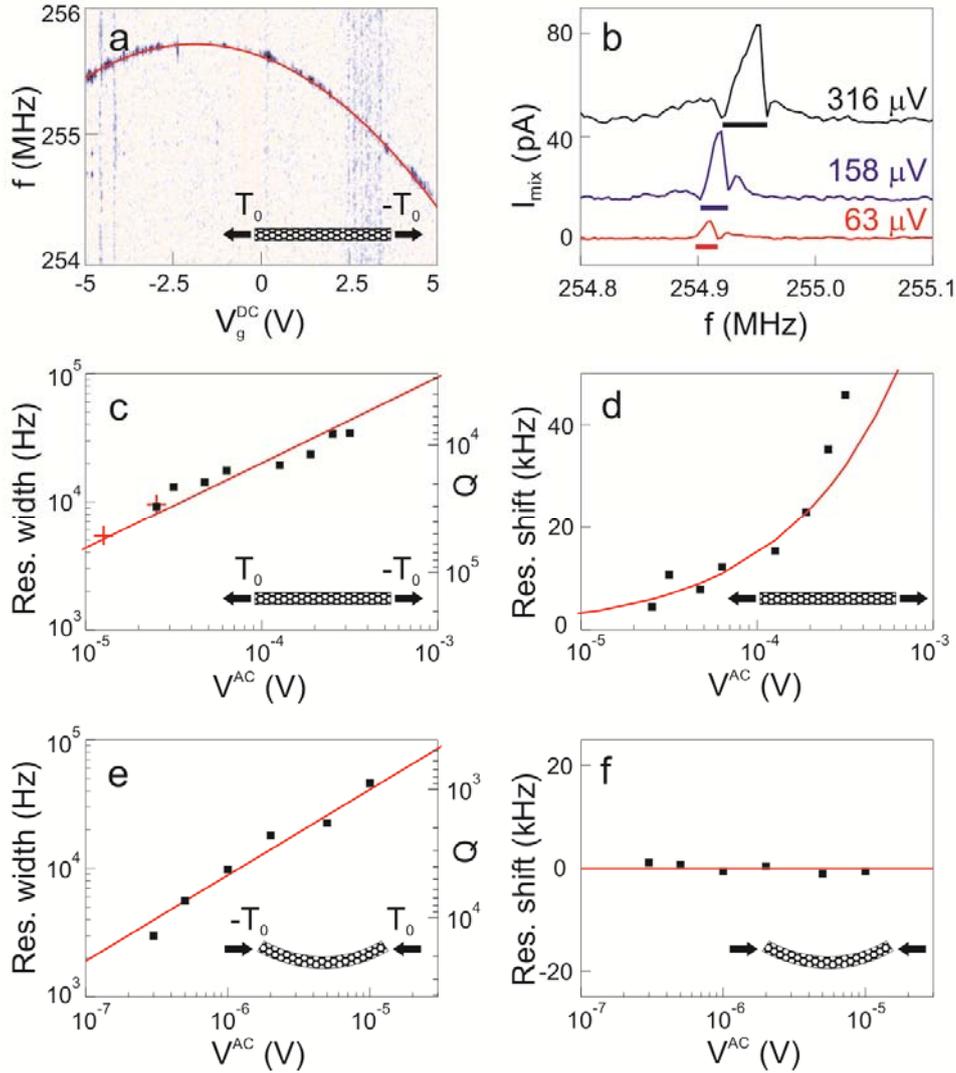

**Figure 2 | Nonlinear damping in nanotube resonators. a,** Resonance frequency as a function of gate voltage $V_g^{DC}$ (by measuring $I_{mix}$ as a function of $f$ and $V_g^{DC}$). The red line corresponds to a fit to a model based on electrostatic considerations (see Supplementary Information). The length of the nanotube is 840 nm and the radius 2 nm. The device is fabricated by depositing the contact electrodes onto the nanotube. **b,** Frequency response of the mixing current for three different driving forces. Both resonance width and resonance frequency become larger when increasing the driving force ($V^{AC}$'s are indicated in the figure). Curves are offset for clarity. **c,** Resonance width as a function of $V^{AC}$. Black squares correspond to 5 K and red crosses to 400 mK. The red line is a solution to equation (2) with $\eta = 10^4$ kg·m$^{-2}$s$^{-1}$ ($\gamma$=0). Q is shown

on the right-hand side scale. **d,** Resonance shift as a function of $V^{AC}$. The red line is a solution to equation (1) with $\alpha = 6 \cdot 10^{12}$ kg·m$^{-2}$s$^{-2}$. **e,, f,** Same as in **c,, d,** but for a nanotube with slack at 100 mK. No resonance shift is apparent. $\eta = 7.9 \cdot 10^{5}$ kg·m$^{-2}$s$^{-1}$ and $\alpha \leq 4.8 \cdot 10^{12}$ kg·m$^{-2}$s$^{-2}$ (the upper limit of $\alpha$ is obtained from the measurement uncertainties). The length of the nanotube is 2 μm and the radius 1.5 nm. The device is fabricated by growing the tube over the predefined electrodes.

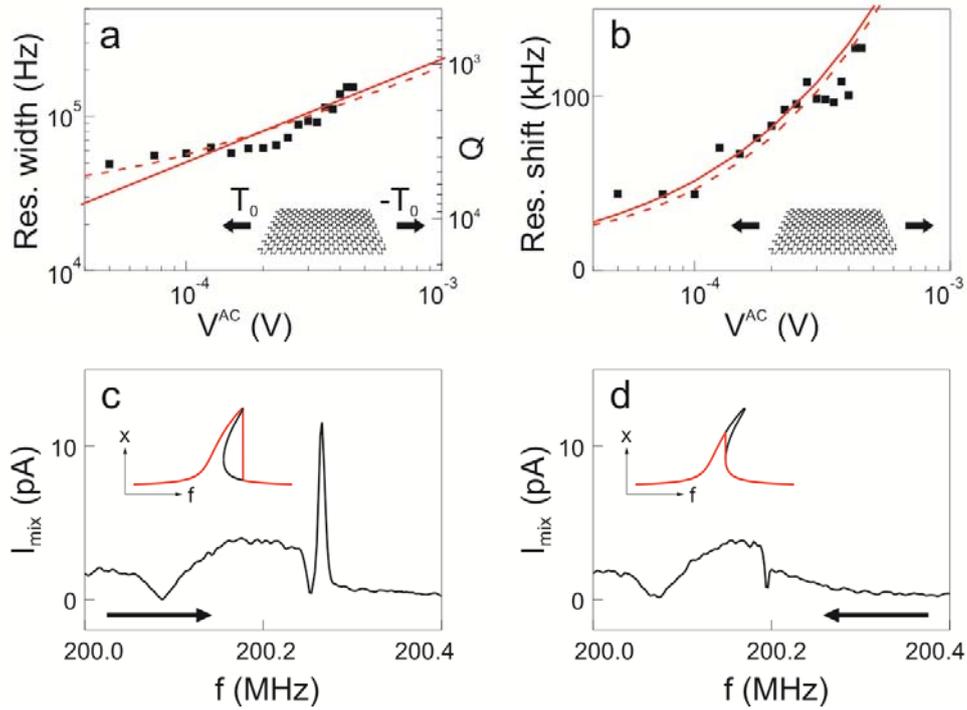

**Figure 3 | Nonlinear damping in a graphene resonator. a, b,** Resonance width and shift as a function of $V^{AC}$ for a graphene sheet under tensile stress at 4 K. The solid red lines are obtained with $\eta = 2.4 \cdot 10^7$ kg·m$^{-2}$s$^{-1}$, $\alpha = 1.9 \cdot 10^{16}$ kg·m$^{-2}$s$^{-2}$ ($\gamma$=0). The dashed red lines represent an improved fit with finite linear damping ($\eta = 1.5 \cdot 10^7$ kg·m$^{-2}$s$^{-1}$, $\alpha = 1.4 \cdot 10^{16}$ kg·m$^{-2}$s$^{-2}$, $\gamma = 8.7 \cdot 10^{-14}$ kg·s$^{-1}$). The length of the graphene is 1.7 μm and the width 120 nm. **c** and **d** show the frequency response of the mixing current at $V^{AC} = 0.5 mV$. The frequency is swept upwards in **c** and downwards in **d**. The schematics in insets show the amplitude of motion as a function of driving frequency (red curves).

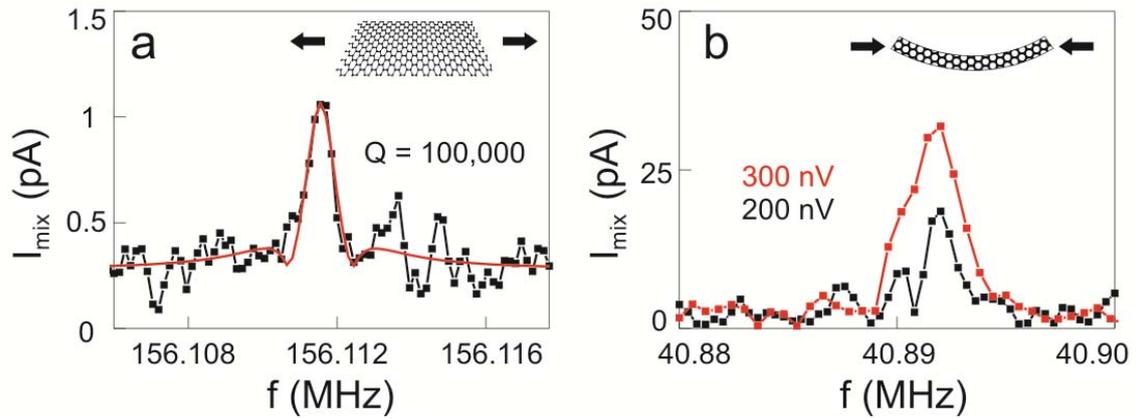

**Figure 4 | Quality factor and force sensitivity at low driving force. a,** Ultra-high quality factor (100,000) for a graphene resonator at 90 mK ($V^{AC} = 8\mu V$). The graphene is under tensile stress and is different from the one in Fig. 3. The red curve is the fit. The length of the graphene is 2 μm and the width 800 nm. Additional resonance curves are shown in Fig. S5. **b,** Ultra-low force sensitivity ($2.5 aN/Hz^{1/2}$) obtained at 100 mK with the nanotube with slack (the same as in Fig.2). The applied $V^{AC}$'s are indicated in the figure.

Supplementary Information

# Nonlinear damping in mechanical resonators based on graphene and carbon nanotubes

A. Eichler[1], J. Moser[1], J. Chaste[1], M. Zdrojek[1], I. Wilson-Rae[2], A. Bachtold[1]

[1] CIN2(ICN-CSIC), Catalan Institute of Nanotechnology, Campus de la UAB 08193 Bellaterra (Barcelona) Spain

[2]Technische Universität München, 85748 Garching, Germany

**A) Device fabrication**

In this work, we present data from three different types of resonators: A nanotube under tensile stress, a nanotube with slack, and a graphene sheet under tensile stress. The fabrication techniques used for each of those devices are different.

**Nanotube under tensile stress**. We grow nanotubes on highly doped, thermally oxidized Si wafers using the chemical vapor deposition (CVD) method [1]. Individual nanotubes are selected with atomic force microscopy (AFM) and localized relative to predefined Au markers. These nanotubes are connected to Cr/Au leads with standard electron-beam lithography (EBL), followed by a thermal evaporation step. Finally, part of the $SiO_2$ underneath the nanotube is etched in hydrofluoric acid (HF) in order to mechanically release the device.

**Nanotube with slack**. We pattern the gate electrode in a trench etched in a highly resistive Si wafer coated with $SiO_2$ and $Si_3N_4$. We then fabricate two W/Pt electrodes that are separated by the trench. We deposit islands of catalyst particles on one of these two electrodes and grow carbon nanotubes by CVD. Many devices are fabricated on the wafer and we choose those for which an electrical contact is established between the contact electrodes. The device is inspected using scanning electron microscopy after the measurements (Fig. 1b) [2-4].

**Graphene sheet under tensile stress**. Graphene flakes are deposited on highly doped, thermally oxidized Si wafers using the adhesive tape technique [5]. Single-layer graphene sheets are selected with an optical microscope by measuring the reflected light intensity using the blue channel of a charged-coupled device camera, the intensity being calibrated with graphene flakes whose number of layers was measured with Raman spectroscopy. The flakes are then cleaned at 300 ºC in an argon/hydrogen atmosphere. In a first EBL/evaporation step, Au markers are added close to the selected flakes, and the localization is repeated more precisely with AFM. The shape of the graphene flakes is tailored in a second EBL step followed by a reactive ion etching process in oxygen. Cr/Au leads are patterned in another EBL step. The graphene sheets are then mechanically released by etching part of the $SiO_2$ in HF [6, 7]. In order to avoid the collapse of the sheets after wet etching due to capillary forces, the devices are successively transferred to water, acetone, dichloroethane, and acetone, and dried in a critical point drier. Before mounting the wafers in the dilution refrigerator, they are annealed in argon/hydrogen at 200 ºC.

**B) Measurement setup**

We perform measurements in a Microkelvin 50-100 dilution refrigerator from Leiden Cryogenics. The radio frequency (RF) signal is transmitted to the source lead (S) through a high frequency coaxial cable with 20 dB attenuators, one at 1 K and another one at 100 mK, and a superconducting coaxial cable between 1 K and 100 mK. For the gate voltage $V_g^{DC}$ and the mixing current, we use shielded manganine wires thermally anchored at the 1 K pot and the mixing chamber. Copper powder filters and RC filters are used close to the sample to attenuate high frequency noise (see Fig. S1). Our RF source is an Agilent E4422B, and the mixing current is measured with a Stanford Research Systems SR830 DSP lock-in amplifier.

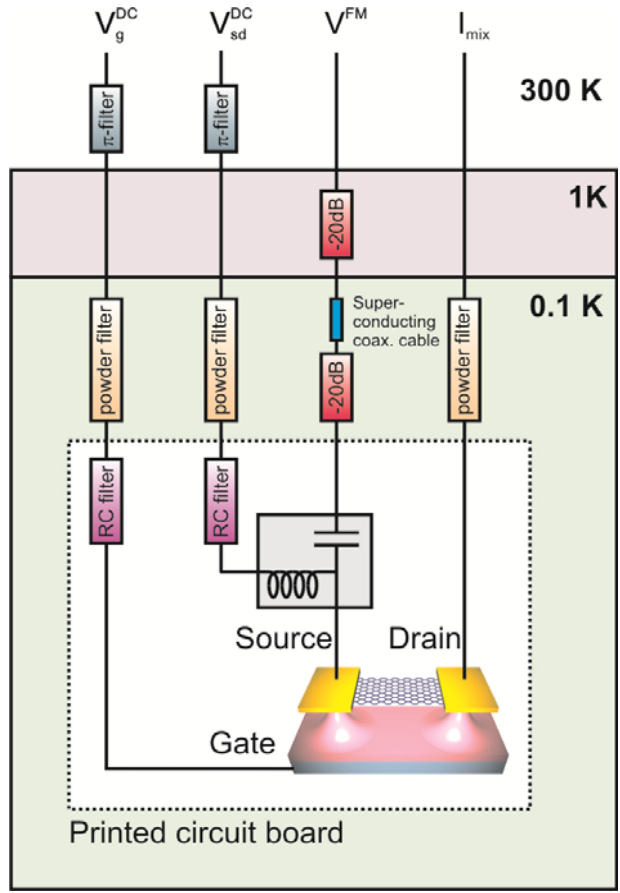

**Figure S1.** Measurement setup schematic showing attenuators and filters at the different stages of the dilution refrigerator. The RF line features a superconducting cable between 100 mK and 1 K.

**C) Frequency modulation mixing technique**

The signal we apply to the source electrode has the form

$$V^{FM}(t) = V^{AC}\cos(2\pi f t + (f_\Delta/f_L)\sin(2\pi f_L t)), \qquad (S1)$$

where $f$ is the carrier frequency, $f_\Delta$ the frequency deviation, $t$ the time, and $f_L$ a low frequency, typically 671 Hz. The resulting mixing current is given by

$$I_{mix} = \frac{1}{2} \cdot \frac{dG}{dV_g^{DC}} \cdot V_g^{DC} \cdot \frac{C'}{C} \cdot V^{AC} \cdot f_\Delta \cdot \frac{\partial}{\partial f} \text{Re}[x_0] \qquad (S2)$$

with $G$ the conductance of the device and $\text{Re}[x_0]$ the real part of its oscillation amplitude [8]. We measure the module of the mixing current with a lock-in amplifier at frequency $f_L$, so our measurement yields $I_{mix} \propto \left| \frac{\partial}{\partial f} \text{Re}[x_0] \right|$. We can see from equation (S2) that there is no purely electrical term in the mixing current in contrast to the more traditional 2-source technique [9]. In addition, we have experienced as in Ref. [8] that the FM technique produces less noise than the 2-source technique.

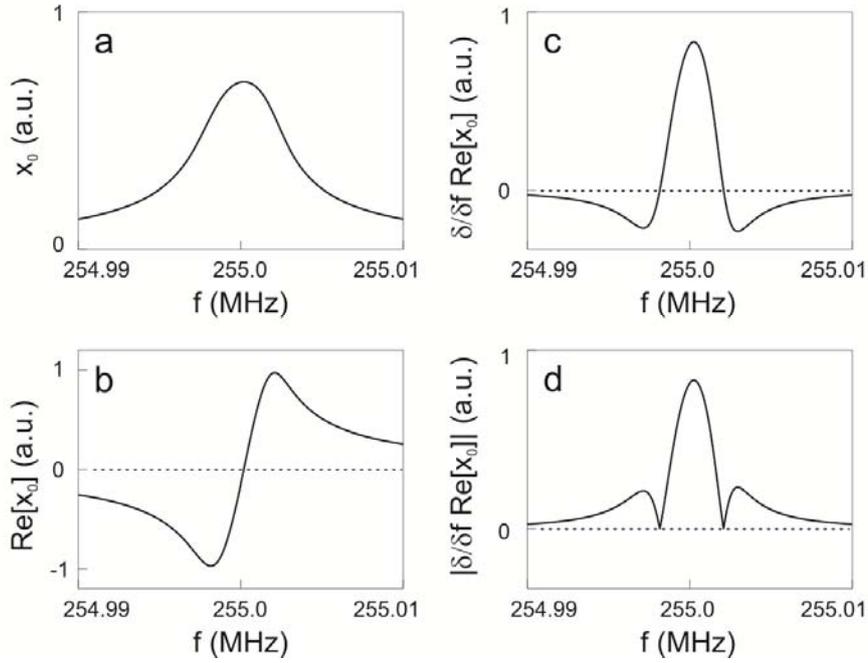

**Figure S2.** Calculated examples of **a** the motion amplitude $x_0$, **b** its real part $\text{Re}[x_0]$, **c** $\frac{\partial}{\partial f} \text{Re}[x_0]$, and **d** $\left| \frac{\partial}{\partial f} \text{Re}[x_0] \right|$.

Fig. S2 depicts calculated examples of the motion amplitude $x_0$, its real part $\text{Re}[x_0]$, $\frac{\partial}{\partial f} \text{Re}[x_0]$, and $\left| \frac{\partial}{\partial f} \text{Re}[x_0] \right|$. The two minima in $\left| \frac{\partial}{\partial f} \text{Re}[x_0] \right|$ (Fig. S2d) provide a

precise and simple way to extract the resonance width, and therefore the quality factor (see section **D**).

Choosing the right value for the frequency deviation ($f_\Delta$) of the FM technique is crucial to a reliable measurement. One has to ensure that $f_\Delta$ is sufficiently small compared to the width of the mechanical resonance $\Delta f$. Otherwise, the measured resonance broadens because the frequency range probed by the FM driving force is too large [8]. In practice, we look for the lowest amplitude of $V^{AC}$ for which we get a reproducible signal, and measure the dependence of $\Delta f$ on $f_\Delta$ (Fig. S3). We select a value for $f_\Delta$ in the plateau at low frequency for which $\Delta f$ corresponds to the dissipation in the resonator (and not to an extrinsic effect related to the measurement). We measure $\Delta f$ at larger $V^{AC}$ keeping the same value for $f_\Delta$.

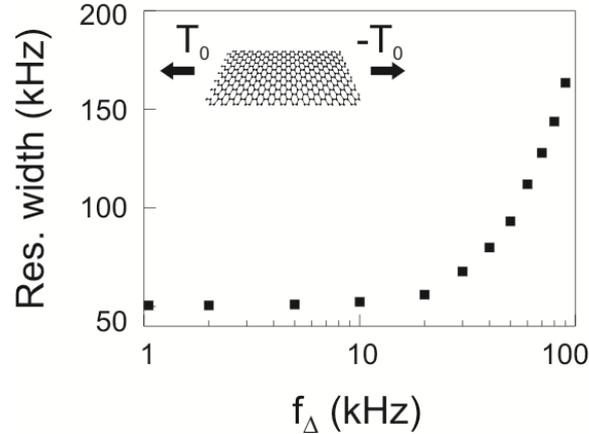

**Figure S3.** Typical dependence of the resonance width on the frequency deviation $f_\Delta$ (for a graphene resonator). The resonance width becomes larger when $f_\Delta > 10$ kHz. The resonance width measured at low $f_\Delta$ is used to extract the quality factor of the resonator.

The reproducibility of the resonance measurements varies from one device to the next. While the resonance lineshape can be very stable in time for some devices (e.g. Fig. S4), fluctuations in $f_0$ and $\Delta f$ can be more pronounced for others. In the latter case, our procedure is to repeat every sweep at least 5 times and take the mean values of $f_0$ and $\Delta f$ (this is especially important for resonances at very low driving voltages where the signal is weak). For example, in the nanotube of Fig. 2c,d fluctuations in $f_0$ and $\Delta f$ are rather large, but they remain lower than the measured broadening and shift of the resonance; the standard deviation of $\Delta f$ is between 0.5 and 2 kHz (depending on $V^{AC}$) and the standard deviation of $f_0$ is around 2 kHz. Figure S5 shows the reproducibility of the resonance of the graphene resonator with $Q = 10^5$.

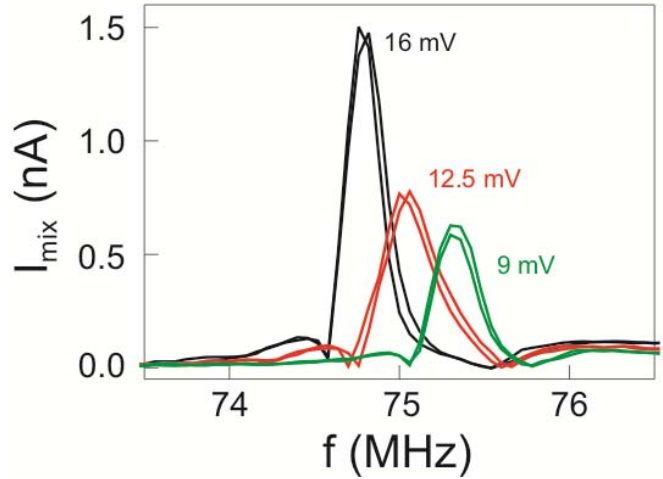

**Figure S4.** Frequency response of the mixing current for different drives at room temperature (same device as in Fig. S10).

In order to discriminate between linear and nonlinear damping, stringent conditions on the measurements have to be met. Indeed, one needs to measure the quality factor by varying the driving force by a large amount (more than one order of magnitude) and keeping $V^{AC}$ below $k_B T / e$ in order to prevent electronic nonlinear effects or local heating. In the present work, we can fulfil such conditions by carrying out the measurements at very low temperature (where the high transconductance allows measuring resonances down to very low $V^{AC}$). In doing so, we can show that the damping is nonlinear. In previous works, the noise in the signal prohibited a systematic study of the nature of the damping. Because there was no obvious reason to expect that nanotube/graphene resonators would behave differently from other resonators, linear damping was tacitly assumed when interpreting previous measurements.

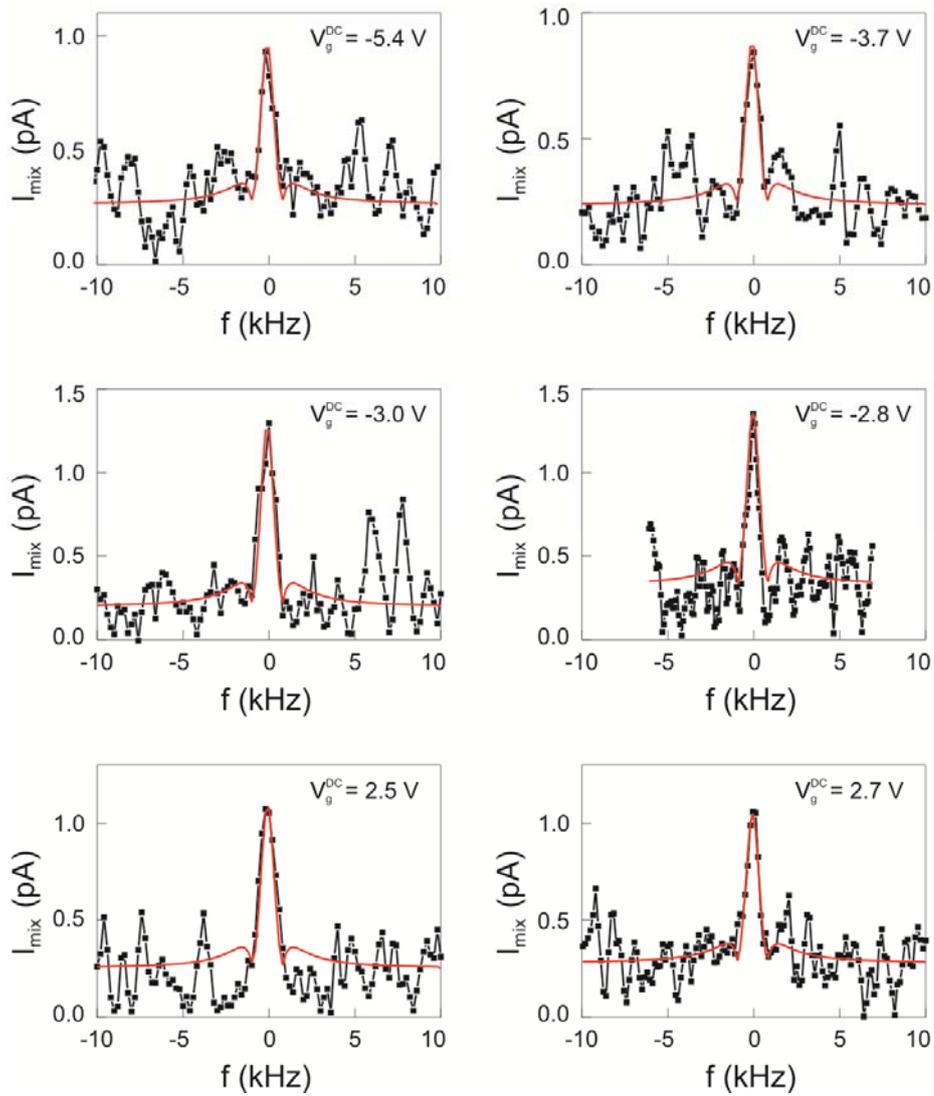

**Figure S5.** Frequency response of the mixing current for different gate voltages (same device as in Fig. 4a). The red curves are fits with $Q = 10^5$. Measurements are carried out at 90 mK with $V^{AC} = 8\ \mu V$. The integration time of the lock-in amplifier is 300 ms.

$f_\Delta = 1$ kHz and similar results are obtained with $f_\Delta = 500$ Hz. The resonance width is typically 1.5 kHz (and occasionally 300 Hz smaller or larger).

**D) Resonance width and quality factor with the FM technique**

**Resonance width:** The mixing current as a function of driving frequency $f$ has a characteristic resonance line-shape with the resonance peak at frequency $f_*$ flanked by two minima. For a weakly damped linear resonator the separation between the latter, $\Delta f$, coincides with the mechanical bandwidth defined as FWHM for the squared modulus of the motional amplitude (time-averaged mechanical energy stored) and allows to extract the quality factor $Q$ in a simple manner from the relation $\Delta f = f_*/Q$ (with $f_* = f_0$). In the presence of nonlinear damping these simple relations break down even in the regime where the response is stable. Nonetheless, here we show that in the limit in which nonlinear damping dominates, i.e. for $2\pi \cdot \Delta f \gg \gamma/m$, analogous relations hold with amplitude independent prefactors close to unity so that (provided that the response is stable) $\Delta f$ still furnishes an adequate measure of the resonance width and allows to determine $Q$ directly (see equation (3)). Similarly, the maximum of the mixing current as a function of $f$, which we use to infer the resonant frequency, and the maximum of the stored energy will be attained for comparable frequencies. The latter, in the relevant case of weak damping and weak anharmonicity (see below), still occurs when $f$ matches the frequency for free undamped oscillations $f_*$, that will now depend on the amplitude.

We follow Ref. [15] and transform equation (1) into a scaled Duffing equation by using the dimensionless variables $\tilde{t} = 2\pi \cdot f_0 t$, $\tilde{x} = \frac{x}{2\pi f_0}\sqrt{\alpha/m}$, and dividing by an overall factor. The corresponding dimensionless parameters read: $1/Q_0 = \frac{\gamma}{2\pi \cdot f_0 m}$, $\tilde{\eta} = 2\pi \cdot f_0 \eta / \alpha$, $\tilde{F}_{drive} = F_{drive} / (2\pi \cdot f_0)^3 \sqrt{\alpha/m^3}$, and $\tilde{f} = f/f_0$. We consider the regime of small oscillations $\tilde{x} \sim \sqrt{\varepsilon} \ll 1$ (i.e. weak damping and weak anharmonicity) via the ansatz $1/Q_0 = \varepsilon$, $\tilde{F}_{drive} = \varepsilon^{3/2} g$, and $\tilde{f} = 1 + \varepsilon\Omega/8$. Note that this treatment also captures the limit where the linear damping term is negligible compared with the cubic one. Secular perturbation theory in $\varepsilon$ allows to determine the steady state complex amplitude at the drive frequency, $\tilde{x}_0 = \sqrt{\varepsilon P} e^{i\phi}$, with $P$ and $\phi$ given by

$$P = \frac{16 g^2}{(\Omega - 3P)^2 + (4 + \tilde{\eta} P)^2} \quad \text{(S3a)}$$

$$\tan\phi = \frac{4 + \tilde{\eta} P}{\Omega - 3P} \quad \text{(S3b)}$$

which are equivalent to equations (S20a) and (S20b) in **H)**.

We define $\Delta f$ as the resonance width inferred from the distance between the minima of the mixing current $I_{mix} \propto \left|\partial/\partial f \, \text{Re}[x_0]\right|$. Thus, in terms of the dimensionless quantities we have $\Delta f = \frac{f_0 (\Omega_+ - \Omega_-)}{8 Q_0}$, where $\Omega_\pm$ are the zeroes of $\partial/\partial\Omega \, \text{Re}[\tilde{x}_0]$. It is straightforward to realize that a sufficient condition for a zero is $F'(\Omega) = 0$, $F(\Omega) \neq 0$ with $F(\Omega) \equiv 16 g^2 \varepsilon^{-1} \text{Re}^2[\tilde{x}_0(\Omega)]$.

Henceforth we focus on the limit $\tilde{\eta} P / 4 \gg 1$ in which the linear contribution to the dissipation is negligible. Thus, we can approximate $4 + \tilde{\eta} P \approx \tilde{\eta} P$ in equations (S3a) and (S3b), which leads to

$$F(\Omega) = \frac{16g^2 P(\Omega)[\Omega - 3P(\Omega)]^2}{(9+\tilde{\eta}^2)P^2(\Omega) - 6\Omega P(\Omega) + \Omega^2} \tag{S4}$$

where $P(\Omega)$ is an implicit function given by

$$(9+\tilde{\eta}^2)P^3 - 6\Omega P^2 + \Omega^2 P - 16g^2 = 0. \tag{S5}$$

We use equation (S5) to simplify the denominator of equation (S4) and arrive at

$$F(\Omega) = P^2(\Omega)[\Omega - 3P(\Omega)]^2, \tag{S6}$$

which combined with the derivative of equation (S5) and the condition $F'(\Omega) = 0$ yields

$$2P'P(\Omega - 3P)(\Omega - 6P) + 2P^2(\Omega - 3P) = 0 \tag{S7}$$

$$P'[3(9+\tilde{\eta}^2)P^2 - 12\Omega P + \Omega^2] + 2\Omega P - 6P^2 = 0. \tag{S8}$$

Subsequently, eliminating $P'$, we obtain

$$(3\tilde{\eta}^2 - 9)P^2 + 6P\Omega - \Omega^2 = 0, \tag{S9}$$

which together with equation (S5) allows us to determine $P(\Omega_\pm) = (2g/\tilde{\eta})^{2/3}$. Finally, substituting this constant into equation (S5) and solving the resulting quadratic equation for $\Omega$ we obtain $\Omega_+ - \Omega_- = \sqrt{3}(32\tilde{\eta}g^2)^{1/3}$, which implies equation (2). The latter can also be expressed as

$$\Delta f = \frac{\sqrt{3}\eta x_*^2}{2^{11/3}\pi m} \tag{S10}$$

in terms of $x_* = \max\{|x_0|\} \propto \max\{\sqrt{P}\}$ (see equation (S11a)). From equation (S3a) we have that the maximum, $P_*$, occurs at $\Omega_* = 3P_*$ corresponding to a driving frequency $f_* = f_0 + \frac{3\alpha x_*^2}{32\pi^2 f_0 m}$ which coincides with the frequency for undamped free oscillations with squared amplitude $x_*^2$ as determined by secular perturbation theory. Thus we find

$$P_* = \left(\frac{4g}{\tilde{\eta}}\right)^{2/3} \tag{S11a}$$

$$x_* = \left(\frac{4F_{drive}}{2\pi \cdot f_0 \eta}\right)^{1/3} \tag{S11b}$$

valid for $\tilde{\eta} P/4 \gg 1$.

Finally, it is natural to compare $\Delta f$ with the FWHM, $\Delta f_{FW}$, of the stored mechanical energy as a function of driving frequency. The latter is proportional to $P(\Omega)$, which for strong nonlinear damping attains "half-maximum" values ($P_*/2$) at $\Omega_{FW}^\pm = P_*(3 \pm \sqrt{7}\tilde{\eta})/2$, as can be verified by direct substitution into equation (S5) (see equation (S11a)). Thus, we obtain

$$\Delta f_{FW} = \frac{\sqrt{7}\eta x_*^2}{32\pi m} \tag{S12}$$

which together with equation (S10) implies the relation

$$\Delta f = 1.65 \Delta f_{FW}. \tag{S13}$$

**Quality factor:** One should note that the nonlinear effects unveiled, though strong when compared with the linear dissipation, still induce frequency shifts and broadenings of the mechanical resonances which are much smaller than the resonant frequency. Therefore the standard definition of $Q$ in terms of the free-ringdown is still meaningful,

$$\frac{1}{Q} = \frac{\langle \dot{E} \rangle}{2\pi f_* \langle E \rangle}, \tag{S14}$$

albeit with an amplitude dependent outcome. Here $E$ is the mechanical energy at a given time and $\langle ... \rangle$ denotes time-averaging over a timescale long compared with the oscillation period but sufficiently short that the decay of the amplitude is negligible. Within the aforementioned approximation scheme (relevant to our scenario) one should consider the denominator to zeroth order in $\varepsilon$ so that $\langle E \rangle \approx 2\pi^2 m f_0^2 |x_0|^2$ and $f_* = f_0$ (here $x_0$ is the amplitude of the free oscillations). In turn, to lowest order in $\varepsilon$ equation (1) yields $\langle \dot{E} \rangle = 2\pi^2 \gamma f_0^2 |x_0|^2 + \pi^2 \eta f_0^2 |x_0|^4 / 2$, which substituted into equation (S14) reduces in the limit $\eta |x_0|^2 / 4 \gg \gamma$ to

$$Q = \frac{8\pi f_0 m}{\eta |x_0|^2}. \tag{S15}$$

Finally, equations (S10) and (S15) imply the relation

$$\left.\frac{1}{Q}\right|_{|x_0|=x_*} = \frac{2^{2/3} \Delta f}{\sqrt{3} f_0} \tag{S16}$$

which is equivalent to equation (3).

**E) Electrical and mechanical characterization of the samples discussed in main text**

**Nanotube under tensile stress (Fig. 2a-d):** Fig. S6 shows the conductance $G$ as a function of gate voltage $V_g^{DC}$ at different temperatures. The conductance is approaching $2 e^2/h$. In addition, the average conductance is increasing upon lowering the temperature. From these two observations, we conclude that the device is in the Fabry-Perot regime [10, 11] and not in the Coulomb blockade regime – namely, the oscillations of the conductance in Fig. S6 are due to quantum electronic interference effects.

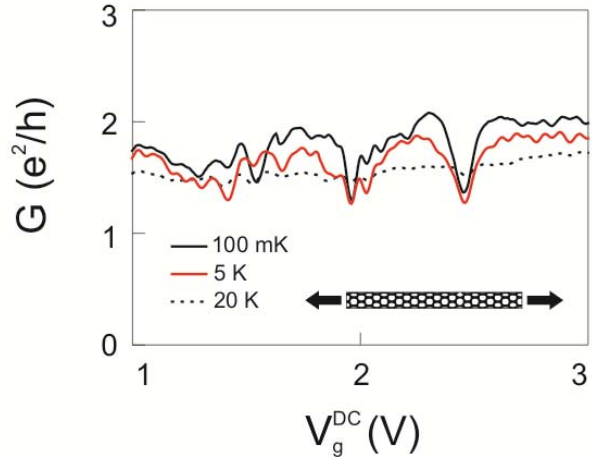

**Figure S6.** Conductance versus gate voltage for the nanotube under tensile stress (Fig. 2a-d) at different temperatures. The device operates in the Fabry-Perot regime.

The $V_g^{DC}$-dependence of $f_0$ (Fig. 2a) is well described by

$$f_0(V_g^{DC}) = f_{max} - \sigma V_g^{DC\,2}, \qquad (S17)$$

where $f_{max} = \frac{1}{2} \cdot \sqrt{T_0/mL}$ and $\sigma = f_{max} C''L/(4\pi^2 T_0)$. The term $f_{max}$ corresponds to the resonance frequency of a beam with built-in tension $T_0$. The term $-\sigma V_g^{DC\,2}$ originates from the oscillation of the beam while a static voltage difference is applied between the beam and the gate (this adds a force $0.5 C''(V_g^{DC})^2 x$ in the equation of motion). From a fit to equation (S17), we obtain $T_0 = 1.7$ nN and $m = 7.87 \cdot 10^{-21}$ kg. This mass is identical to the one expected for a nanotube with radius $r = 2$ nm and length $L = 840$ nm ($r$ and $L$ are measured with AFM).

We calculate the capacitance and its differentiations with respect to the distance between the nanotube and the gate $\xi$ using

$$C = \frac{2\pi\varepsilon_0 L}{\ln(2\xi/r)}, \qquad (S18)$$

with $\xi = 440$ nm. We obtain $C = 7.67 \cdot 10^{-18}$ F, $C' = -2.87 \cdot 10^{-12}$ F/m, and $C'' = 8.5 \cdot 10^{-6}$ F/m².

The Duffing nonlinearity $\alpha x^3$ can have a geometrical origin [15]. Namely, it can arise from stretching upon deflection of the resonator, a consequence of the clamping at both ends. In the high tension regime $T_0 \gg ESr^2/L^2$ of elastic thin rod theory applied to a hollow tube, we find $\alpha = \pi^4 ES/4L^3$ for the fundamental resonance. Here, $T_0$ is the built-in tension, $E = 1$ TPa the Young's modulus, $S$ the cross section, $r$ the radius, and $L$ the length. Using $S = \pi r^2 - \pi(r - 0.16nm)^2$ with the measured $r = 2$ nm, we obtain $\alpha = 7.5 \cdot 10^{13}$ kg·m⁻²s⁻² for the nanotube under tensile stress in Fig. 2. This value compares favorably with the one obtained from the fit ($\alpha = 6 \cdot 10^{12}$ kg·m⁻²s⁻²); the agreement can be improved by using a larger value for $L$. This is sound since clamping is not perfect, i.e. the finite rigidity of the metal electrodes implies that the vibrations extend into the region of the contacts; in addition, the nanotube may slide underneath the electrodes. As for the other resonators discussed in the main text, it is difficult to estimate $\alpha$ in a reliable way since the eigenmodes of a nanotube with slack are nontrivial [12] and the width to length ratio for graphene resonators is not small enough to warrant the use of thin rod elasticity. Another Duffing nonlinearity can stem from electrostatic effects [15], yet these are negligible since the estimated $\alpha$ is almost four orders of magnitude lower than what we measure (see section **I**).

**Nanotube with slack (Fig. 2e-f and Fig. 4b):** In Fig. S7a, we plot the mixing current as a function of driving frequency *f* and gate voltage $V_g^{DC}$. Since the resonances are difficult to see in this figure, they are highlighted with lines in Fig. S7b. The gate voltage dependence of the lowest mode is linear and we detect multiple eigenmodes. These are two signatures of a suspended nanotube resonator with slack [9, 12]. We note that the mechanical oscillations are not detected for $V_g^{DC} > 0$ because the transconductance is much lower.

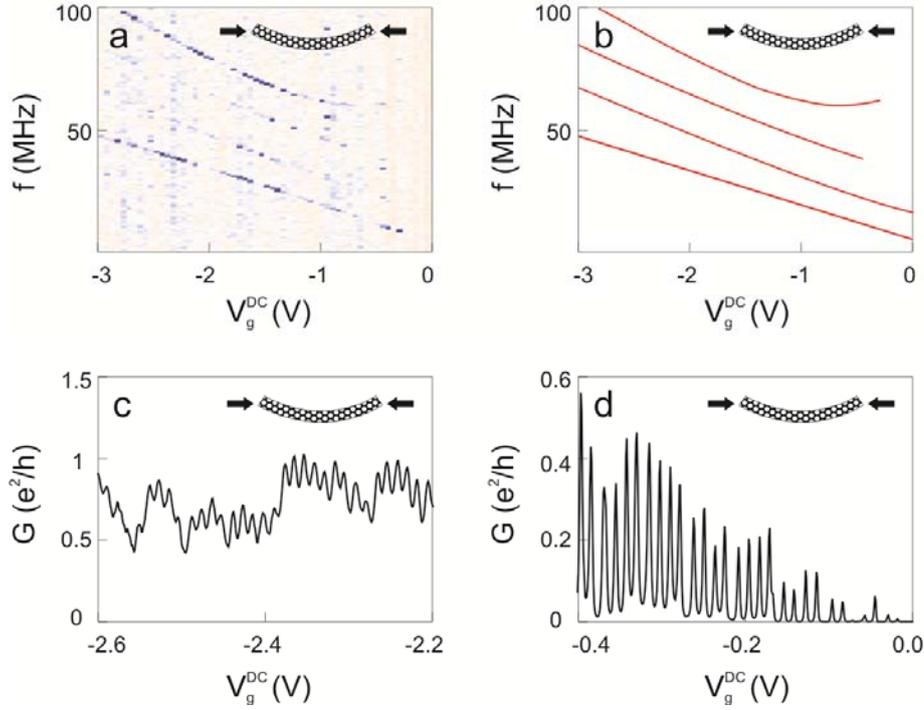

**Figure S7. a,** Mixing current as a function of frequency and gate voltage at 100 mK for the nanotube with slack (Fig. 2e-f). Large current appears dark blue. **b,** Schematic of the $V_g^{DC}$ dependence of the resonance frequencies for better visibility. **c** and **d,** Conductance as a function of the gate voltage for different $V_g^{DC}$ regions.

G versus $V_g^{DC}$ is shown in Fig. S7c and Fig. S7d for two different regimes of Coulomb blockade. In Fig. S7c the Coulomb blockade is in the regime of strong coupling between the nanotube and the contact electrodes. The conductance is close to e²/h and the oscillations are low in amplitude (the temperature is 100 mK and the charging energy is about 1.5 meV).The data in Fig. 2e,f (main text) are measured in this regime ($V_g^{DC} = -2.3$ V). Figure S7d shows the regime of Coulomb blockade in the weak coupling limit. The conductance goes to zero between the Coulomb blockade peaks.

The gate-nanotube capacitance $C = 11.9$ aF is estimated using the simple relation $C = e/\Delta V_g^{DC}$, where e is the elementary charge and $\Delta V_g^{DC}$ the separation between two Coulomb blockade peaks. We estimate $C' = -5.2 \cdot 10^{-12}$ F/m from

$$C' = -\frac{C}{\xi \ln(2\xi/r)} \tag{S19}$$

after having determined $\xi = 370$ nm and $r = 1.5$ nm by AFM (the radius is measured on the portion in contact with the metal electrodes).

Since this nanotube is grown in the last step of the fabrication process (see section **A**), contamination should be low. We estimate $m = 1.4 \cdot 10^{-20}$ kg from the radius $r = 1.5$ nm and the length $L = 2$ μm (the length is measured with scanning electron microscopy after the measurements).

**Graphene sheet under tensile stress:** Fig. S8a displays the mixing current for the same graphene sheet as in Fig. 3 (main text) as a function of driving frequency *f* and gate voltage $V_g^{DC}$. The sheet length is $1.7$ μm and the width $120$ nm. We perform a fit to the $V_g^{DC}$ dependence of the resonance frequency $f_0$ in the same way as above for the nanotube under tensile stress, (red parabola). From the fit, we infer an estimation of the mass ($m = 3.9 \cdot 10^{-19}$ kg) and the built-in tensile stress ($T_0 = 110$ nN). The mass of a pristine graphene sheet with the given geometry is $m = 1.6 \cdot 10^{-19}$ kg. The difference might originate from contamination [7].

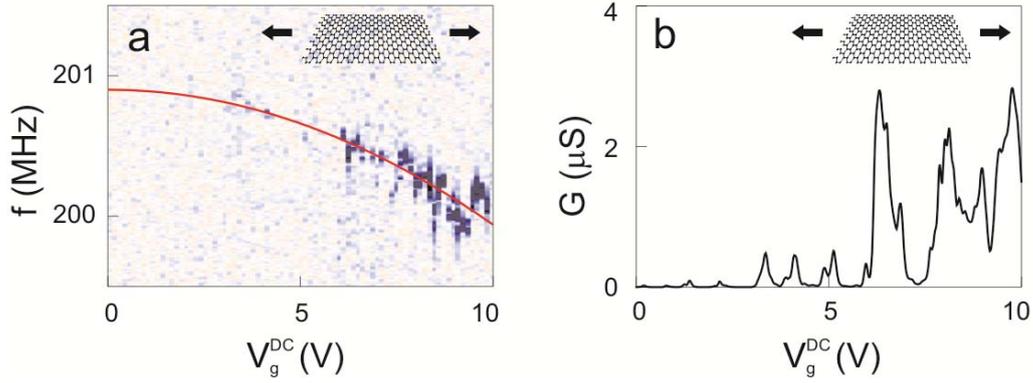

**Figure S8. a,** Mixing current as a function of frequency and gate voltage for the same graphene resonator as in Fig. 3. Large current appears dark blue. **b,** The conductance shows strong modulations as a function of gate voltage. The temperature is 2K.

Fig. S8b reveals the variation of the conductance with gate voltage. Measurements in Fig. 3 of the main text are recorded at $V_g^{DC} = 7.8$ V, a $V_g^{DC}$ region where the modulation of G is attributed to strong localization possibly in combination with charging effects [13].

**F) Additional graphene device (not shown in the main text)**

A fourth set of data is summarized in Fig. S9. The width and length are both $1.3$ μm. The plot of mixing current versus $f$ and $V_g^{DC}$ allows us to extract the built-in tensile stress ($T_0 = 5.2 \cdot 10^{-7}$ N) and the mass ($m = 1.3 \cdot 10^{-18}$ kg) (Fig. S9a). The mass is identical to the one expected from the size of the sample.

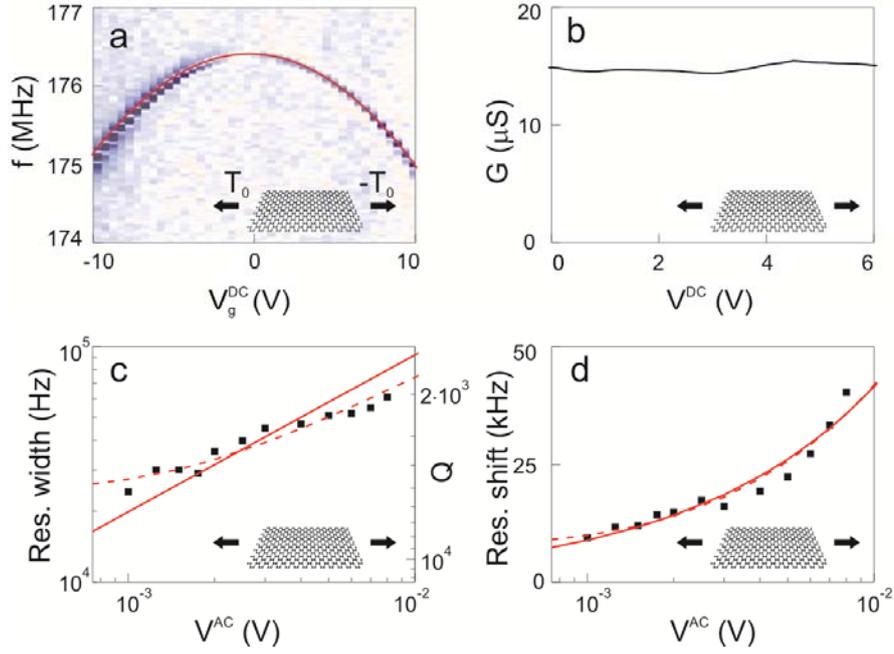

**Figure S9. a,** Mixing current as a function of frequency and gate voltage for a graphene resonator under tensile stress that is not discussed in the main text. Large current appears dark blue. **b,** The conductance shows only little modulation as a function of gate voltage. **c,** Resonance width and **d,** resonance shift as a function of $V^{AC}$. Solid red lines represent a comparison to equation (2) with negligible linear damping ($\eta = 1.6 \cdot 10^4$ kgm$^{-2}$s$^{-1}$, $\alpha = 3.1 \cdot 10^{12}$ kgm$^{-2}$s$^{-2}$), dashed lines are obtained with finite $\gamma$ ($\eta = 5.9 \cdot 10^3$ kgm$^{-2}$s$^{-1}$, $\alpha = 1.74 \cdot 10^{12}$ kgm$^{-2}$s$^{-2}$, $\gamma = 2.0 \cdot 10^{-13}$ kgs$^{-1}$). The temperature is 2 K.

The conductance shows only minor modulations as a function of $V_g^{DC}$ (Fig. S9b). These modulations are attributed to universal conductance fluctuations [14], which originate from quantum electronic interference effects.

Fig. S9c and Fig. S9d show the $V^{AC}$ dependence of the resonance width and the resonance shift, respectively. Measurements are taken at $V_g^{DC} = 4$ V. Setting $\gamma$ to zero (negligible linear damping), we obtain $\eta = 1.6 \cdot 10^{4}$ kg·m$^{-2}$s$^{-1}$ and $\alpha = 3.1 \cdot 10^{12}$ kg·m$^{-2}$s$^{-2}$ (solid red lines). The fit to the resonance width can be improved by introducing finite linear damping (dashed red lines, $\eta = 5.9 \cdot 10^{3}$ kg·m$^{-2}$s$^{-1}$, $\alpha = 1.74 \cdot 10^{12}$ kg·m$^{-2}$s$^{-2}$, $\gamma = 2.0 \cdot 10^{-13}$ kg·s$^{-1}$). The corresponding ratio $\eta \cdot 2\pi \cdot f_0 / \alpha = 3$ is in agreement with the fact that we observe no hysteresis in this resonator.

**G) Additional nanotube device measured at 300 K (not shown in main text)**

We present measurements taken at 300 K on an additional nanotube resonator in Fig. S10. The nanotube is grown in the last fabrication step like the one in Fig. 2e-f of the main text.

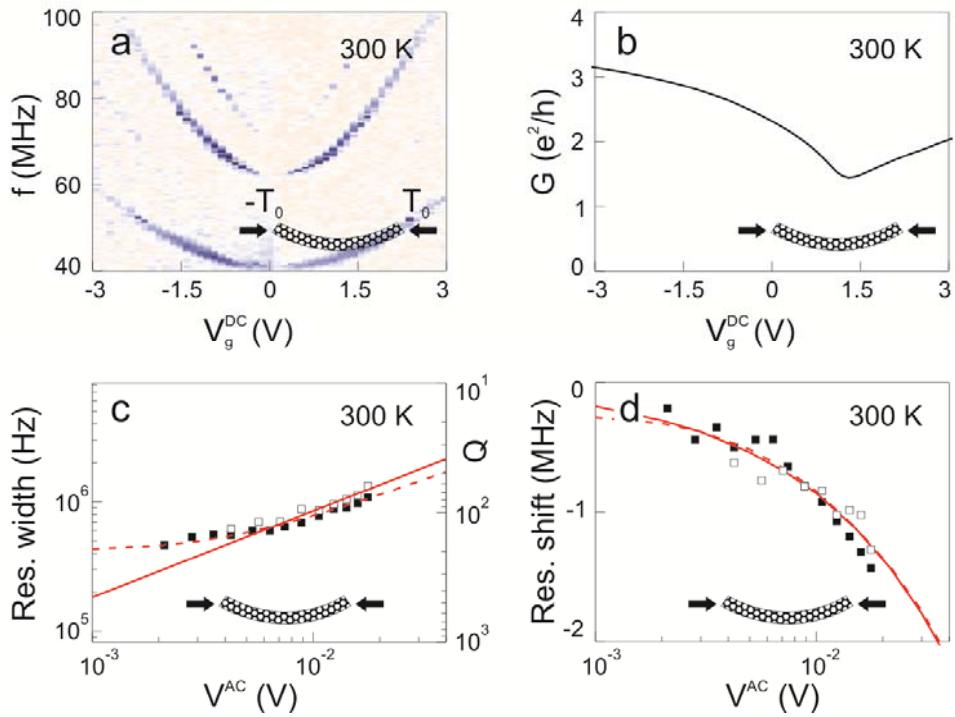

**Figure S10. a,** Mixing current as a function of frequency and gate voltage at 300 K. Large current appears dark blue. **b,** Conductance versus gate voltage at 300 K. **c,** Resonance width and **d,** resonance shift measured with the FM technique (black squares) and 2-source technique (hollow squares), respectively. Solid red lines represent a comparison to equation (2) with negligible linear damping ($\eta = 2.5 \cdot 10^{3}$

kg·m$^{-2}$s$^{-1}$, $\alpha = -7 \cdot 10^{11}$ kg·m$^{-2}$s$^{-2}$, $\gamma = 0$), dashed lines are obtained with a finite $\gamma$ ($\eta = 1 \cdot 10^3$ kg·m$^{-2}$s$^{-1}$, $\alpha = -4.4 \cdot 10^{11}$ kg·m$^{-2}$s$^{-2}$, $\gamma = 1.9 \cdot 10^{-14}$ kg·s$^{-1}$).

Fig. S10a shows the mixing current versus $f$ and $V_g^{DC}$. We observe multiple eigenmodes with concave parabolic shape, suggesting that the resonator has slack.

The conductance of the device rises close to 4 e$^2$/h (Fig. S10b). Measurements of the mechanical characteristics are performed for $V_g^{DC} = 1.7$ V, where the modulation of the conductance with $V_g^{DC}$ is comparably large.

Black squares in Fig. S10c and d show the broadening and the shift of the resonance upon increasing $V^{AC}$. The resonance frequency shifts towards lower values, in contrast to the other devices we present.

With best estimates for the device geometry ($\xi = 370$ nm, $r = 1.5$ nm, $L = 1$ μm) we calculate $m = 7 \cdot 10^{-21}$ kg and $C' = -4 \cdot 10^{-12}$ F/m. In turn, these values enable us to perform a fit to the data assuming negligible linear damping ($\eta = 2.5 \cdot 10^3$ kg·m$^{-2}$s$^{-1}$, $\alpha = -7 \cdot 10^{11}$ kg·m$^{-2}$s$^{-2}$, $\gamma = 0$, solid red lines). The fit to the resonance width is much improved when taking into account finite linear damping ($\eta = 1 \cdot 10^3$ kg·m$^{-2}$s$^{-1}$, $\alpha = -4.4 \cdot 10^{11}$ kg·m$^{-2}$s$^{-2}$, $\gamma = 1.9 \cdot 10^{-14}$ kg·s$^{-1}$, dashed red lines). The corresponding ratio $\eta \cdot 2\pi \cdot f_0 / \alpha < \sqrt{3}$ is in agreement with the fact that we observe a hysteresis in this resonator.

We repeat the experiment, this time employing the 2-source mixing technique instead of the FM technique to measure the mechanical resonance. We find good agreement between the two methods (hollow and filled squares in Fig. S10c-d).

**H) Fitting procedure of the $V^{AC}$ dependence of the resonance width and the resonance shift**

In this section, we will describe the fitting procedure used in Fig. 2c-f, Fig. 3a-b, Fig. S9c-d, and Fig. S10c-d. Equation (1) of the main text leads to

$$x_0^2 = \frac{(F_{drive}/4\pi f_0)^2}{\left(m(2\pi f - 2\pi f_0) - \frac{3}{16}\frac{\alpha}{\pi f_0}x_0^2\right)^2 + \left(\frac{1}{2}\gamma + \frac{1}{8}\eta x_0^2\right)^2} \quad \text{(S20a)}$$

$$\tan(\phi) = \frac{\gamma/2 + \eta x_0^2/8}{m(2\pi f - 2\pi f_0) - \frac{3}{16}\frac{\alpha}{\pi f_0}x_0^2} \quad \text{(S20b)}$$

for the motion amplitude $x_0$ and the phase $\phi$ [15]. Solving these expressions numerically, we calculate $x_0$, $\text{Re}[x_0] = x_0 \cos(\phi)$, and finally $I_{mix} \propto \left|\partial/\partial f \, \text{Re}[x_0]\right|$ (see Fig. S2a-d). Here $x_0$ is the maximum amplitude, which is attained at the midpoint (i.e. the mechanical eigenmode $u(z)$ satisfies $u(L/2)=1$), and the parameters in equation (1) are scaled so that $m$ corresponds to the total suspended mass rather than the effective mass of the mode. We note that this naturally affects the expression for the $\alpha$ expected from the geometric nonlinearity.

The determination of $\eta$ and $\alpha$ takes place in several steps. We first assume that $\gamma = 0$, which implies that the resonance width scales as $\Delta f \propto (V^{AC})^{2/3}$. We extract $\eta$ using $\Delta f = 0.032 m^{-1} \eta^{1/3} f_0^{-2/3} F_{drive}^{2/3}$. We then perform a fit of the resonance shift as a function of the driving force $F_{drive} = gC'V_g^{DC}V^{AC}$ by solving equation (S20a) and (S20b) with the Duffing term $\alpha$ as free parameter. The parameter *g* accounts for the shape of the eigenmode (for example, $g = 4/\pi$ for a beam under tensile stress). More precisely, we assume a constant load (valid within the model used for the capacitances) which together with the aforementioned normalization and parameter scaling leads to

$$g = \frac{\int_0^L u(z)dz}{\int_0^L u(z)^2 dz} > 1. \qquad (S21)$$

Heuristic considerations imply that for the fundamental mode *g* is always of order unity. For measurements where the resonance width tends to saturate at low $V^{AC}$, the fit can be improved using a finite $\gamma$. In this case, we perform a fit of the resonance width and the resonance shift as functions of $F_{drive} = gC'V_g^{DC}V^{AC}$ by solving equation (S20a) and (S20b) with $\gamma$, $\alpha$, and $\eta$ as free parameters.

In order to quantify the shift of the resonance frequency both for the experimental data and the calculations, we use the frequency where the mixing current has its maximum.

**I) Electrostatic Duffing nonlinearities**

Electrostatic nonlinearities arise when applying a voltage difference between an oscillating beam and a nearby gate electrode. The electrostatic force reads $F_{electrostatic} = 0.5 C'(\xi)V_g^{DC^2}$ where $C'(\xi)$ depends on the beam motion. We get

$$F_{electrostatic} = 0.5 V_g^{DC^2}(C' + C''\delta\xi + C'''\delta\xi^2 + C''''\delta\xi^3) \qquad (S22)$$

assuming small motion amplitude $\delta\xi$. The Duffing term is thus $\alpha_{el} = \frac{1}{2}C''''V_g^{DC^2}$ and it is negative ($C'''' < 0$). This results in a softening of the linear spring constant, in contrast to our experimental findings in Fig. 2d or Fig. 3b.

We estimate the electrostatic Duffing term of the nanotube resonator under tensile stress by calculating $C''''$ from equation (S18). We estimate that $C'''' = 3.57 \cdot 10^8$ F/m$^4$ and $\alpha_{el} = -1.1 \cdot 10^9$ kg·m$^{-2}$s$^{-2}$, more than three orders of magnitude lower than the fitted value.

We also evaluate the electrostatic Duffing term of the graphene sheet in Fig. 3. Using the capacitive plate model we have $C'''' = 24\varepsilon_0 A / \xi^5$, where $\varepsilon_0 = 8.85 \cdot 10^{-12}$ F/m, $A$ the area of the suspended sheet, and $\xi$ the sheet-gate separation. We get $\alpha_{el} = -4.3 \cdot 10^{11}$ kg·m$^{-2}$s$^{-2}$, which is almost five orders of magnitude smaller than the fitted value. In summary, the electrostatic Duffing nonlinearities are not relevant to our experiments.

**J) The broadening of the resonance width is not associated to the coupling between electrons and mechanical vibrations**

The coupling between electrons and mechanical vibrations can be very strong in nanotubes and can lead to important nonlinearities [16,17]. However, the broadening of the resonance width discussed in this work is not associated to the electron-vibration coupling.

We first note that the electron-vibration coupling is only strong when the transport is in the Coulomb blockade regime. However, the nanotube in Fig. 2c is in the Fabry-Perot regime, the graphene sheets in Fig. 4a, S5, and S9 are deeply in the diffusive regime, and the nanotube in Fig. S10 is measured at room temperature. To be more specific, we estimate the associated damping for the nanotube measured at room temperature in Fig. S10c. Using the relation

$$1/Q = \frac{C'^2}{2\pi f_0 m} \frac{V_g^{DC\,2}}{G} \tag{S23}$$

valid when the transport is not in Coulomb blockade (supplementary information of [16]) with $C' = -4 \cdot 10^{-12}$ F/m, $V_g^{DC} = -1.5$ V, $f_0 = 47$ MHz, $m = 7 \cdot 10^{-21}$ kg, and $G = 9 \cdot 10^{-5}$ S, the quality factor related to the electron-vibration coupling is $5 \cdot 10^6$. This is more than 4 orders of magnitude larger than the $Q$ that we measure, showing that the electron-vibration coupling is weak.

Due to Coulomb blockade, the electron-vibration coupling can become nonlinear (i.e. the nonlinear coupling is equivalent to an electrostatic force acting on the resonator that is nonlinear in displacement, as discussed in detail in [16]). This effect stems from the Coulomb staircase (the averaged charge of the dot is highly nonlinear with regard to the control charge). However, the electrical transport in our present work is in most cases not Coulomb blockaded and the nonlinearity in the electron-vibration coupling disappears (the averaged charge in the device is linear in the control charge to a large extent).

Another important point are nonlinearities in the detection. We impose stringent measurement conditions by keeping $V^{AC}$ lower than $k_B T/e$. In this case, the relation between current and voltage remains linear to a very good accuracy.

Overall, because the transport is not in the Coulomb blockade regime in most cases and the excitation is lower than $k_B T/e$, an influence of the electron-vibration coupling on the observed broadening of the resonance width can be ruled out.

1. Lassagne, B., Garcia-Sanchez, D., Aguasca, A., & Bachtold, A. Ultrasensitive Mass sensing with a nanotube electromechanical resonator. *Nano Lett.* **8,** 3735-3738 (2008).

2. Cao, J., Wang, Q., & Dai, H. Electron transport in very clean, as-grown suspended carbon nanotubes. *Nature Mat.* **4,** 745-749 (2005).

3. Hüttel, A. K., Steele, G. A., Witkamp, B., Poot, M., Kouwenhoven, L. P., & van der Zant, H. S. J. Carbon nanotubes as ultrahigh quality factor mechanical resonators. *Nano Letters* **9,** 2547-2552 (2009).

4. Steele, G. A., Götz, G., & Kouwenhoven, L. P. Tunable few-electron double quantum dots and Klein tunnelling in ultraclean carbon nanotubes. *Nature Nanotech.* **4,** 363-367 (2009).

5. Novoselov, K. S., Geim, A. K., Morozov, S. V., Jiang, D., Zhang, Y., Dubonos, S. V., Grigorieva, I. V., & Firsov, A. A. Electric field effect in atomically thin carbon films. *Science* **306,** 666-669 (2004).

6. Bolotin, K. I., Sikes, K. J., Jiang, Z., Fundenberg, G., Hone, J., Kim, P., & Stormer, H. L. Ultrahigh electron mobility in suspended graphene. *Solid State Communications* **146,** 351-355 (2008).

7. Chen, C., Rosenblatt, S., Bolotin, K. I., Kalb, W., Kim, P., Kymissis, I., Stormer, H. L., Heinz, T. F., & Hone, J. Performance of monolayer graphene nanomechanical resonators with electrical readout. *Nature Nanotech.* **4,** 861-867 (2009).

8. Gouttenoire, V., Barois, T., Perisanu, S., Leclercq, J.-L., Purcell, S. T., Vincent, P., & Ayari, A. Digital and FM demodulation of a doubly clamped single-walled carbon-nanotube oscillator: towards a nanotube cell phone. *Small* **6,** 1060-1065 (2010).

9. Sazonova, V., Yaish, Y., Üstünel, H., Roundy, D., Arias, T. A., & McEuen, P. L. A tunable carbon nanotube electromechnaical oscillator. *Nature* **431,** 284-287 (2004).

10. Liang, W., Bockrath, M., Bozovic, D., Hafner, J. H., Tinkham, M., & Park, H. Fabry - Perot interference in a nanotube electron waveguide. *Nature* **411,** 665-669 (2001).

11. Kong, J., Yenilmez, E., Tombler, T. W., Kim, W., Dai, H., Laughlin, R. B., Liu, L., Jayanthi, C. S., & Wu, S.Y. Quantum interference and ballistic transmission in nanotube electron waveguides. *Phys. Rev. Lett.* **87,** 106801 (2001).

12. Üstünel, H., Roundy, D., & Arias, T. A. Modeling a suspended nanotube oscillator. *Nano Letters* **5,** 523-526 (2005).

13. Oostinga, J. B., Sacépé, B., Craciun, M. F., & Morpurgo, A. F. Magnetotransport through graphene nanoribbons. *Phys. Rev. B* **81,** 193408 (2010).

14. Morozov, S. V., Novoselov, K. S., Katsnelson, M. I., Schedin, F., Ponomarenko, L. A., Jiang, D., & Geim, A. K. Strong suppression of weak localization in graphene. *Phys. Rev. Lett.* **97,** 016801 (2006).

15. Lifshitz, R., & Cross, M. C. *Reviews of Nonlinear Dynamics and Complexity* (Wiley-VCH, New York, 2008, Vol. 1)

16. Lassagne, B, Tarakanov, Y., Kinaret, J., Garcia-Sanchez, D., & Bachtold, A. Coupling mechanics to charge transport in carbon nanotube mechanical resonators. *Science* **325,** 1107-1110 (2009).

17. Steele, G. A., Hüttel, A. K., Witkamp, B., Poot, M., Meerwaldt, H. B., Kouwenhoven, L. P., & van der Zant, H. S. J. Strong coupling between single-electron tunneling and nanomechanical motion. *Science* **325,** 1103-1107 (2009).